\documentclass[twocolumn, epjc3]{svjour3}

\RequirePackage[T1]{fontenc}

\smartqed  

\RequirePackage{graphicx}
\RequirePackage{mathptmx}      
\RequirePackage{flushend}
\RequirePackage[colorlinks, citecolor = blue, urlcolor = blue, linkcolor = blue]{hyperref}

\journalname{Eur. Phys. J. C}

\usepackage{amsmath}
\usepackage{mathtools}
\DeclarePairedDelimiter\abs{\lvert}{\rvert}  
\usepackage{color}
\usepackage{epsfig}
\usepackage{multirow}
\usepackage{overpic}
\usepackage{colortbl}
\usepackage{float}
\usepackage{appendix}
\usepackage[labelformat = simple]{subcaption}

\usepackage{textcomp}
\usepackage{booktabs}
\usepackage{siunitx}[=v2]
\DeclareSIUnit\clight{\text{\ensuremath{c}}}
\usepackage{tikz}
\usepackage{upgreek}
\usepackage{bm}
\usepackage{xspace}
\usepackage{verbatim}
\usepackage[backend = biber, style = phys, biblabel = brackets, articletitle = true, eprint]{biblatex}
\begin{document}
\title{Sensitivity study of the charged lepton flavor violating process $\tau \to \gamma \mu$ at STCF}
\author{%
  Teng Xiang\thanksref{e1,addr1,addr2}
  \and
  Xiao-Dong Shi\thanksref{addr3,addr4}
  \and
  Da-Yong Wang\thanksref{addr1,addr2}
  \and
  Xiao-Rong Zhou\thanksref{addr3,addr4}
}
\thankstext{e1}{\email{xiangteng@pku.edu.cn}}
\institute{%
  Peking University, Beijing 100871, People's Republic of China\label{addr1}
  \and
  State Key Laboratory of Nuclear Physics and Technology, Beijing 100871, People's Republic of China\label{addr2}
  \and
  University of Science and Technology of China, Hefei 230026, People's Republic of China\label{addr3}
  \and
  State Key Laboratory of Particle Detection and Electronics, Hefei 230026, People's Republic of China\label{addr4}
}
\date{Received: date / Accepted: date}

\maketitle

\begin{abstract}
  A sensitivity study for the search for the charged lepton flavor violating process $\tau \to \gamma\mu$ at the Super $\tau$-Charm Facility is performed with a fast simulation. With the expected performance of the current detector design and an integrated luminosity of \SI{1}{ab^{-1}} corresponding to one-year of data taking, the sensitivity on the branching fraction (BF) of $\tau \to \gamma\mu$ is estimated to be at the level of \num{e-8}. The sensitivity under different detector performances are also studied. With ideal performance, the BF could be probed to be \num{2.8e-8} at \SI{90}{\percent} confidence level. The sensitivity is expected to scale with the square root of the luminosity, therefore with a total luminosity of \SI{10}{ab^{-1}} corresponding to ten-year of data taking, the sensitivity could reach \num{8.8e-9}, which is about one order of magnitude improvement upon the current best upper limit.
\end{abstract}

\section{Introduction}
In the Standard Model (SM), the charged lepton flavor violating (cLFV) processes can occur through neutrino oscillation, but are highly suppressed due to the small mass of neutrino~\cite{SM_prediction}. For example, the branching fraction (BF) of $\ell_1 \to \ell_2 \gamma$ in the SM is
\begin{equation}
  \text{BF}(\ell_1 \to \ell_2 \gamma) = \frac{3\alpha_e}{32\pi}\left|\sum_i U_{1i}^* U_{2i} \frac{m_i^2}{M_W^2}\right|^2 \approx 10^{-50} \sim 10^{-54},
\end{equation}
where $U$ is the PMNS matrix~\cite{PMNS}, $i$ runs over the three neutrinos, $m_i$ and $m_W$ are the masses of neutrinos and $W$ boson. The BF in the SM is well beyond the sensitivity of current experiments, thus the observation of cLFV processes would be an unambiguous signature of new physics. On the other hand, the lepton flavor conservation, differing from other conservation laws in the SM, is not associated with an underlying conserved current, therefore many theoretical models beyond the SM naturally introduce cLFV processes, such as the Minimal Supersymmetric SM~\cite{MSSM}, the Grand Unified Theories~\cite{GUT1,GUT2}, and seesaw mechanisms~\cite{seesaw}. Some of them predict BFs that are close to the current experimental sensitivity.

As the heaviest lepton, tau has many possible cLFV decay modes, amongst them $\tau \to \gamma\mu$ is regarded as one of the best probes, and is predicted in a wide variety of models with rates enhanced to observable level. For example, the BF is predicted to be up to \num{e-9} in seesaw models~\cite{prediction_seesaw}, \num{e-10} in Higgs-mediated SUSY models~\cite{prediction_MSSM_1,prediction_MSSM_2}, \num{e-8} in SO(10) SUSY models~\cite{prediction_SUSY_SO10_1,prediction_SUSY_SO10_2}, and \num{e-9} in non-universal $Z'$ models~\cite{prediction_Zprime}. Experimentally, the most stringent upper limit (UL) on the BF of this channel is given by BABAR to be \num{4.4e-8} at \SI{90}{\percent} confidence level~(C.L.)~\cite{BABAR} and Belle to be \num{4.2e-8} at \SI{90}{\percent} C.L.~\cite{Belle}. Next generation of experiments are aiming at pushing the sensitivity down for another one order of magnitude or even further~\cite{Snowmass_tau_LFV}.

The proposed Super $\tau$-Charm Facility (STCF)~\cite{STCF_CDR} in China, which is an electron-positron collider with a center-of-mass energy in the region of the $\tau$-charm threshold, is one of such next generation of experiments. In this paper, the sensitivity of searching for $\tau \to \gamma\mu$ at STCF is studied to explore the physics potential of STCF and guide the design of the experiment. STCF has several advantages on searching for $\tau \to \gamma\mu$. As an electron-positron collider, the total four-momentum is known and the final state is fully reconstructed, leading to higher efficiency and lower background. The energy of STCF can be adjusted to be just above the threshold of tau pair production $e^+e^- \to \tau^+\tau^-$, where the cross-section reaches the maximum~\cite{tau_pair_xs}, and the energy of radiative photons which is one of the main sources of fake signal photons in background is low thus can be well separated from signal.

\section{Detector design and Monte Carlo simulation}
The proposed STCF is a symmetric electron-positron collider operating at center-of-mass energy $\sqrt{s}$ from \SI{2.0}{GeV} to \SI{7.0}{GeV} with a designed peaking luminosity over \SI{0.5e35}{cm^{-2}.s^{-1}}~\cite{STCF_CDR}. Such an environment will serve as an important high statistics and low background platform to test the SM and probe possible new physics beyond the SM, such as cLFV decays of tau. Assuming that it runs at $\sqrt{s} = \SI{4.26}{GeV}$, STCF will accumulate \num{3.5e9} tau pairs~\cite{KKMC1,KKMC2} per year for the corresponding expected integrated luminosity of \SI{1}{ab^{-1}}.

As a general purpose detector designed for the electron-positron collider, the STCF detector consists a tracking system composed of the inner and outer trackers, a particle identification (PID) system with charged kaon/pion misidentification rate less than \SI{2}{\percent} up to \SI{2}{GeV/\clight}, an electromagnetic calorimeter (EMC) with an excellent energy resolution and a good time resolution, a super-conducting solenoid, and a muon detector (MUD) that provides good charged pion/muon separation. The design and the expected performances for each sub-detector are detailed in the conceptual design report~\cite{STCF_CDR}.

At present, the STCF detector and the corresponding offline software system are under research and development. A fast simulation software package is therefore developed to access the physics study~\cite{FastSim}, which takes the most common event generators as input to perform a fast and realistic simulation. The simulation includes resolution and efficiency responses for tracking of final state charged particles and photons, PID system and kinematic fit related variables. Instead of performing the detailed simulation of interaction of the final objects with detector by {\sc Geant4}~\cite{Geant4}, the responses of objects in each sub-system, including efficiency, resolution and other variables used in physics analyses, are modeled by sampling the shape and magnitude according to their performances. By default, all the parameters are based on empirical formulae or extracted from simulation of the BESIII detector~\cite{BESIII,BOSS}, while the fast simulation also provides flexible interface for adjusting performance of each sub-system, which can be used to optimize the detector design according to the physics requirements.

This sensitivity study is performed based on Monte Carlo (MC) samples at \SI{4.26}{GeV}, including samples for signal and expected possible background processes. Using the tag method with selected modes as detailed in Sec.~\ref{sec:analysis_procedure}, the major expected background processes are those that are discussed in the rest of this section.  One of them is the dimu process $e^+e^- \to (\gamma)\mu^+\mu^-$ where the two muons and radiated photons can be misidentified as the tag pion and the signal muon and photon. Another background is the ditau process $e^+e^- \to \tau^+\tau^-$ where one tau decay is necessarily the same as tag side of signal event and the other tau decay can be misidentified as $\tau \to \gamma\mu$. The hadronic process $e^+e^- \to q\bar{q} (q = u, d, s, c)$ could also contribute to the background due to its high cross-section at this energy region. Samples for these processes are simulated with high luminosity, corresponding to \SI{1}{ab^{-1}}, \SI{6}{ab^{-1}}, and \SI{1}{ab^{-1}}, respectively. The dimu samples are generated with {\sc Babayaga}~\cite{Babayaga1,Babayaga2}. For the ditau samples, the production of tau pair is generated with {\sc KKMC}~\cite{KKMC1,KKMC2} and the decay of tau with {\sc Tauola}~\cite{Tauola}. Hadronic process is generated with {\sc LundArlw}~\cite{LundArlw}. Other processes at this energy region are expected to be negligible. The background level of Bhabha process $e^+e^- \to e^+e^-$ and digamma process $e^+e^- \to \gamma\gamma$ are much lower than dimu process due to the low probability of electron misidentified as muon or the lack of charged tracks. For the two-photon process $e^+e^- \to e^+e^-X$ where $X$ is any system generated from the virtual photons radiated by the initial electron and positron, the electron and position tend to have small scattering angles, thus escape from the detector along the beam energy and cause large energy loss, and the remaining $X$ can hardly pass the event selection. Dedicated MC samples for these processes are also generated, and it is demonstrated that they can be removed at early stage of event selection. The signal MC samples are simulated as process $e^+ e^- \to \tau^+ \tau^-$ with one tau goes to SM decay modes and the other decays to $\gamma\mu$. The decay $\tau \to \gamma\mu$ is generated with the pure phase space model since the dynamics is unknown.

\section{Analysis procedure}
\label{sec:analysis_procedure}
\begin{figure}[htbp]
  \centering
  \includegraphics[width = \linewidth]{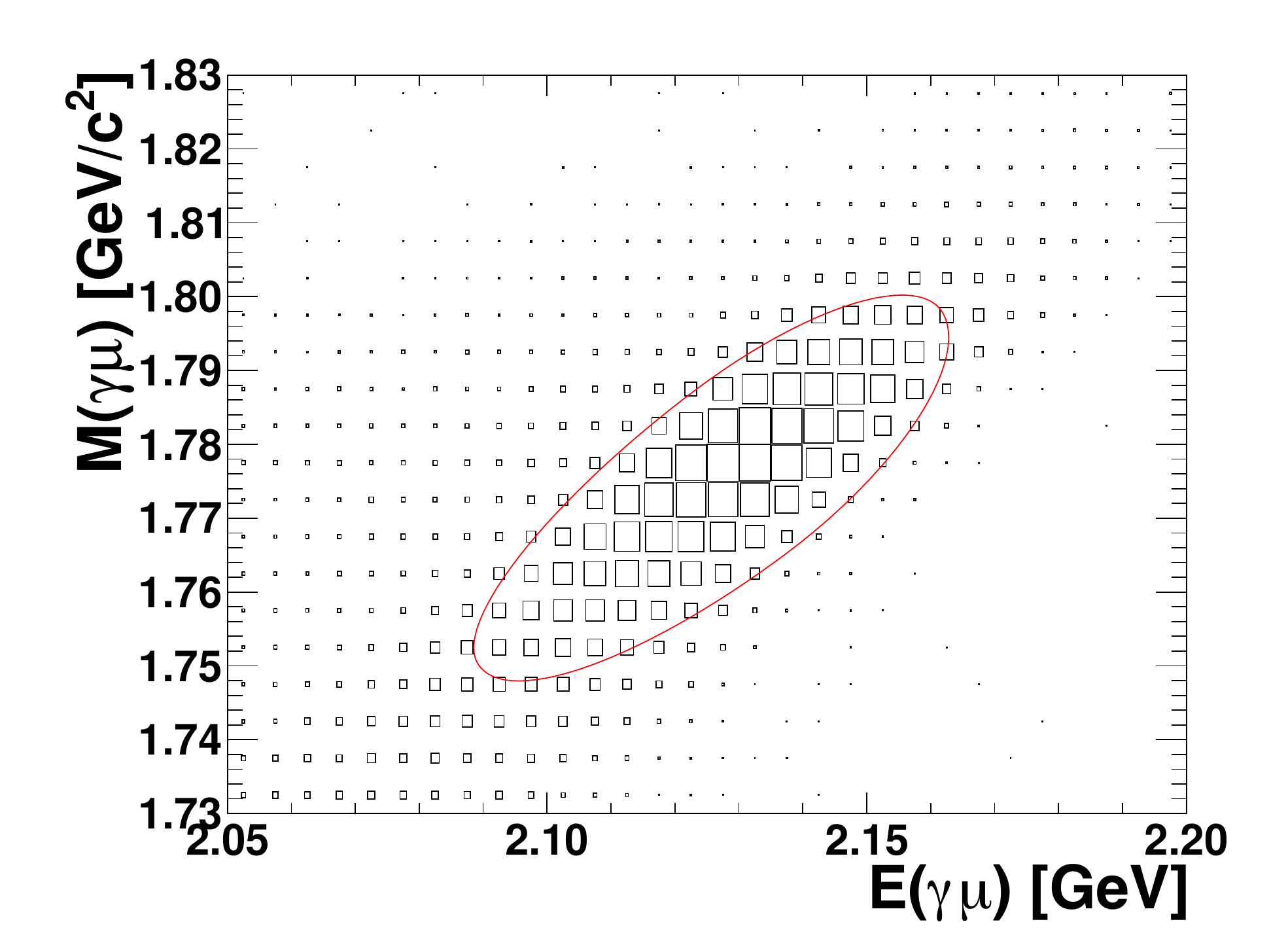}
  \caption{The box plot is the two-dimensional distribution of $E(\gamma\mu)$ and $M(\gamma\mu)$ of signal samples where larger box indicates higher density, and the red line shows the signal region.}
  \label{fig:signal_region_2D}
\end{figure}

At STCF, $\tau^+$ and $\tau^-$ are produced in pairs, so we can tag $\tau^+$ (denoted as tag side) by its SM decay modes and search for cLFV decay of its partner $\tau^-$ (denoted as signal side). The charge-conjugated channels are always implied throughout the paper. For the tag side, amongst the five main 1-prong decay modes of $\tau^+$, $e^+ \nu_e \bar{\nu}_\tau$, $\pi^+ \bar{\nu}_\tau$ and $\pi^+ \pi^0 \bar{\nu}_\tau$ are selected as tag modes, which account for \SI{54}{\percent} of the total BF of tau decays~\cite{PDG}. $\tau^+ \to \mu^+ \nu_\mu \bar{\nu}_\tau$ mode is not used due to high $e^+ e^- \to (\gamma) \mu^+ \mu^-$ background. As for $\tau^+ \to \pi^+ \pi^0 \pi^0 \bar{\nu}_\tau$ mode, due to the high photon multiplicity, the reconstruction efficiency is low and combinatorial background is high, and the tag photons can be easily misidentified as signal photon. The signal side consists of a signal photon and a signal muon and is featured by a peak around the beam energy on the total energy distribution and a peak around the tau mass on the invariant mass distribution of signal photon and muon, as shown in Fig.~\ref{fig:signal_region_2D}.

\begin{figure}[htbp]
  \begin{subfigure}{0.49\linewidth}
    \centering
    \includegraphics[width = \linewidth]{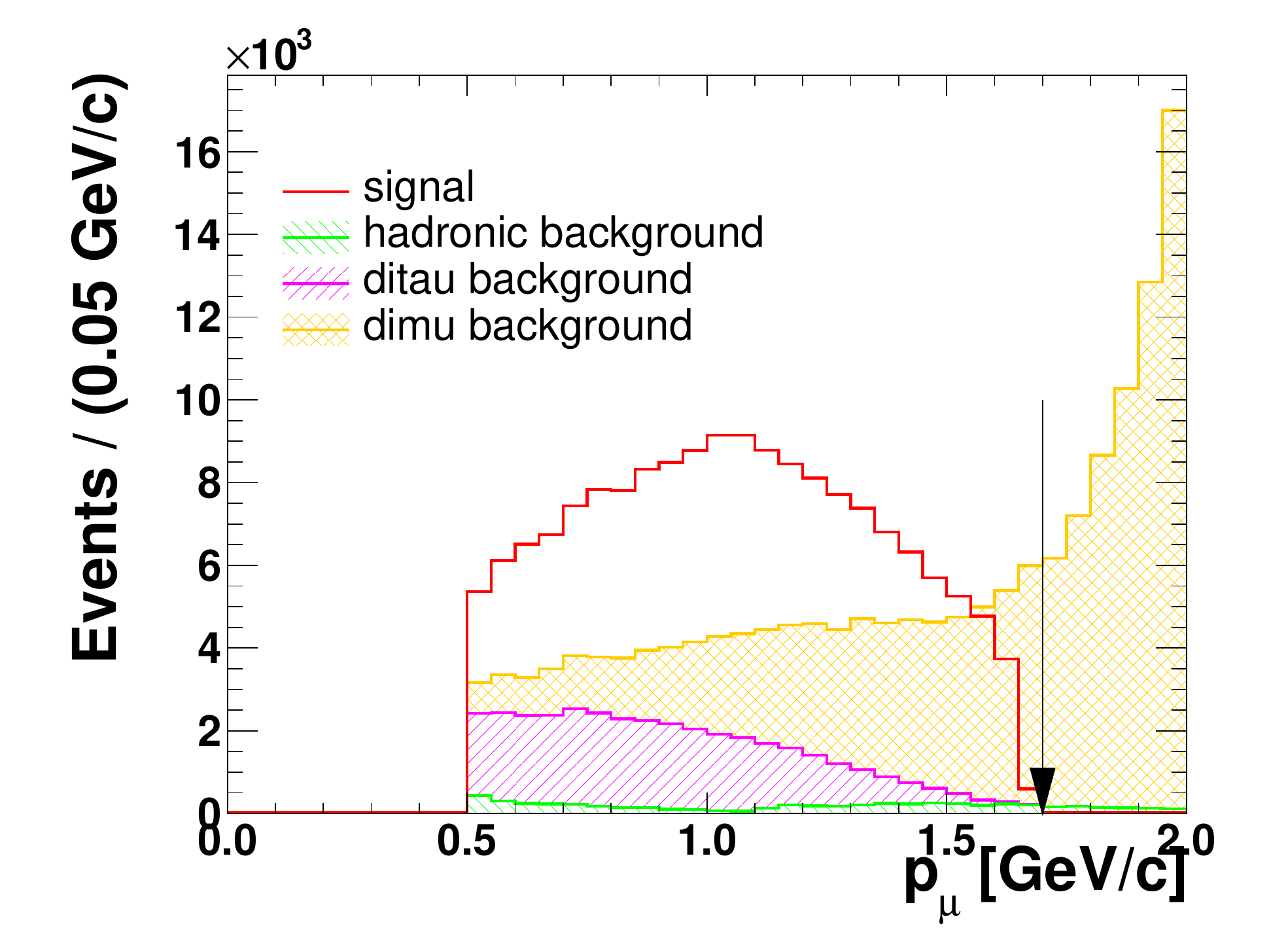}
    \caption{}
    \label{subfig:PHSP_p_mu}
  \end{subfigure}
  \hfill
  \begin{subfigure}{0.49\linewidth}
    \centering
    \includegraphics[width = \linewidth]{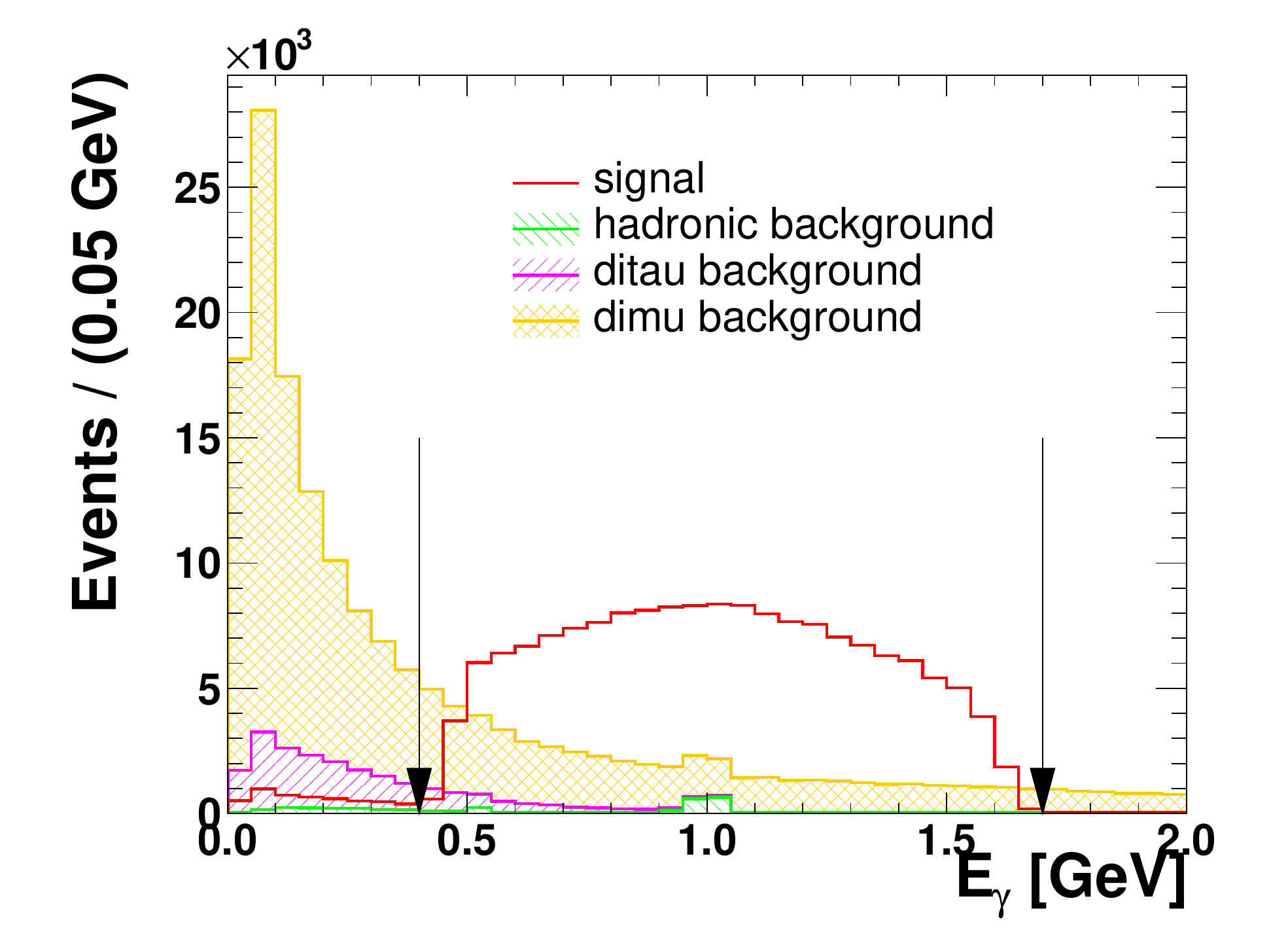}
    \caption{}
    \label{subfig:PHSP_E_gamma}
  \end{subfigure}
  \begin{subfigure}{0.49\linewidth}
    \centering
    \includegraphics[width = \linewidth]{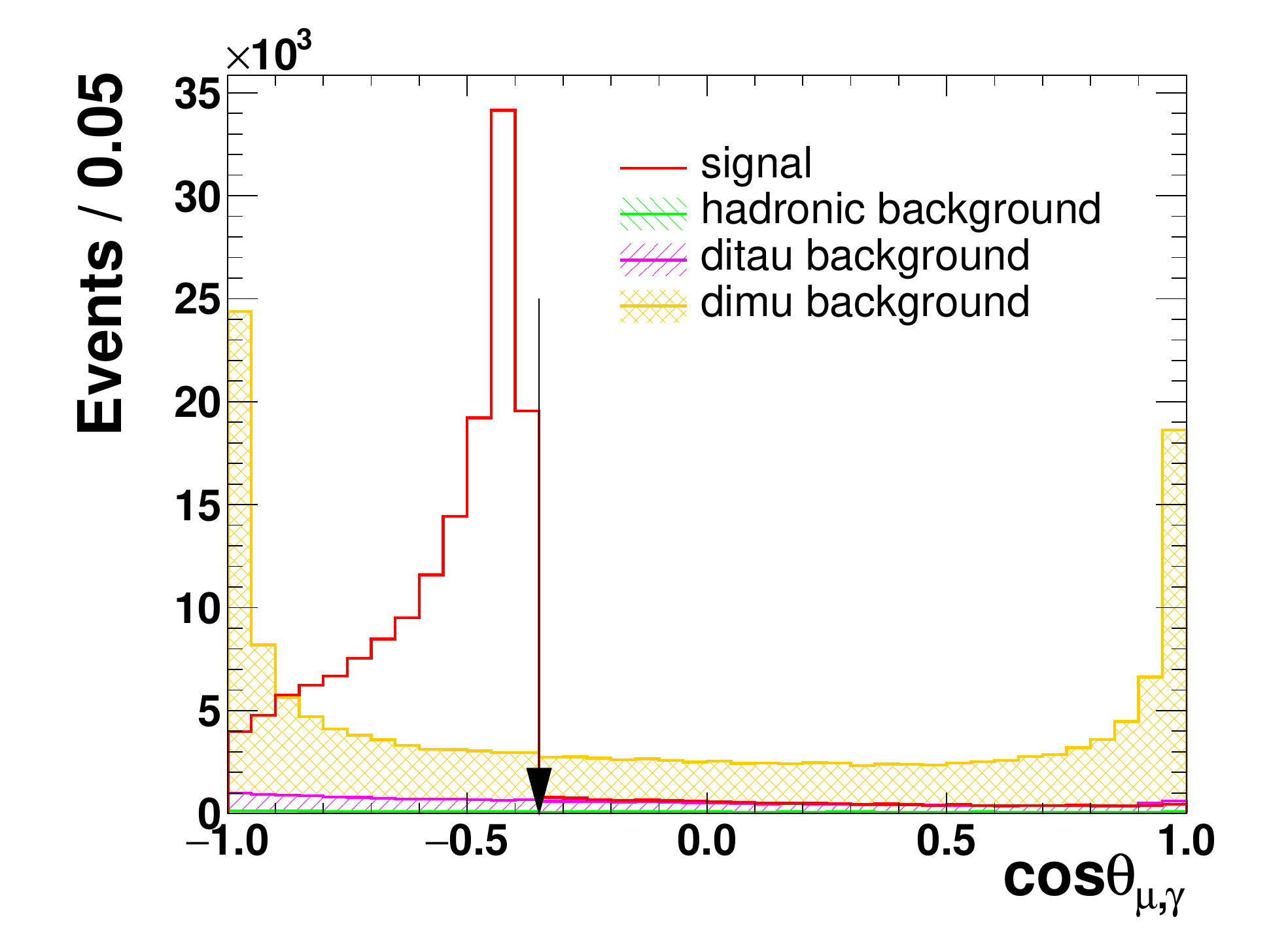}
    \caption{}
    \label{subfig:PHSP_cos_theta_sig_gamma_mu}
  \end{subfigure}
  \caption{The kinematic distributions of \subref{subfig:PHSP_p_mu} momentum of signal muon, \subref{subfig:PHSP_E_gamma} energy of signal photon, and \subref{subfig:PHSP_cos_theta_sig_gamma_mu} cosine of angle between signal photon and muon. The open histogram is signal, and the hatched histograms are background drawn in stacks. The arrows mark the event selection criteria. The lower end cutoff on momentum of muon is caused by MUD acceptance.}
  \label{fig:PHSP_p_mu_cos_theta_gammamu}
\end{figure}

Charged tracks are selected after passing the criteria in fast simulation. For each charged track, the polar angle with respect to beam direction is required to satisfy $\abs{\cos\theta} < 0.93$, and the closest approach to the interaction point must be within \SI{10}{cm} in the beam direction and \SI{1}{cm} in the plane perpendicular to the beam. Events with two charged tracks and zero net-charge are selected. PID is also performed for charged tracks by the fast simulation, where the identification of electron, pion and kaon is based on extracted likelihood of each particle type hypothesis calculated from the information of PID system and the one with the highest likelihood is assigned, while muon is identified by MUD based on preliminary research and development results. One of the tracks is required to be identified as electron or pion, which is tag charged particle, and the other is required to be identified as muon, which is signal muon. $E/p$ information is then used for further identification, where $E$ is the deposited energy in EMC and $p$ is the momentum of the track. $E/p > 0.8$ is required for electron and $E/p < 0.5$ for muon and pion. Photons are selected with an energy threshold of \SI{25}{MeV} in EMC barrel region ($\abs{\cos\theta} < 0.8$) and \SI{50}{MeV} in EMC end cap region ($0.86 < \abs{\cos\theta} < 0.92$) to reject electronic noise and unrelated energy depositions. Neutral pions are reconstructed with two-photon combinations with invariant masses within a $\pi^0$ mass window of around \SI{0.12}{GeV/\clight^2} to \SI{0.14}{GeV/\clight^2} determined by fitting. The exact mass window depends on the detector performance design described later in Sec.~\ref{sec:optimization}. It is required that there is exactly one remaining photon after neutral pions reconstruction and it is denoted as the signal photon. To further suppress background, the momentum of the signal muon and energy of the signal photon are both required to be in \SIrange[range-phrase = {,\ }, range-units = brackets, open-bracket = [, close-bracket= ],]{0.4}{1.7}{GeV}, and the angle between them is required to satisfy $\cos\theta_{\gamma\mu} < -0.35$, all of which are constrained by the kinematics (shown in Fig.~\ref{fig:PHSP_p_mu_cos_theta_gammamu}). Finally, a two-dimensional signal region is chosen on the total energy $E(\gamma\mu)$ and invariant mass $M(\gamma\mu)$ distributions of signal photon and signal muon. Since the two distributions are asymmetric and correlated, the signal region is an asymmetric oblique ellipse, as shown in Fig.~\ref{fig:signal_region_2D}.

The tag mode for each event is then assigned based on the event selection result. The event is classified into $e^+ \nu_e \bar{\nu}_\tau$ mode if one electron is identified, and it is required that there are no neutral pions reconstructed. If one charged pion is identified, the event is further classified into $\pi^+ \bar{\nu}_\tau$ or $\pi^+ \pi^0 \bar{\nu}_\tau$ mode based on whether the number of neutral pions is 0 or 1. Events that do not fit into the classifications are discarded.

After above initial selections, there are mainly five kinds of background classified according to the tag modes and background processes. For $e^+ \nu_e \bar{\nu}_\tau$ tag mode, the main background is ditau process where tag tau radiative decays to electron, signal tau SM decays to muon, and signal photon is misidentified from radiative photon of tag side. For $\pi^+ \bar{\nu}_\tau$ tag mode, the main background is dimu and ditau processes. In the dimu process, one muon is misidentified as tag pion, the other muon is regarded as signal muon, and signal photon is misidentified from radiative photon. For backgrounds from the ditau process, signal tau SM decays to muon, tag tau decays to $\pi\pi^0$ with $\pi^0$ not successfully reconstructed and the daughter photons are regarded as signal photon or not detected. For the $\pi^+ \pi^0 \bar{\nu}_\tau$ tag mode, one of the main background is the ditau process with tag tau decaying to $\pi\pi^0$ or $\pi\pi^0\pi^0$ which is similar to the ditau background process in $\pi^+ \bar{\nu}_\tau$ tag mode. The other is hadronic process, mainly $\pi^+\pi^- + (\geq 1)\pi^0$, where the signal muon is misidentified from charged pion and the signal photon is from $\pi^0$(s) that are not successfully reconstructed.

\begin{figure}[htbp]
  \begin{subfigure}{0.49\linewidth}\centering\includegraphics[width = \linewidth]{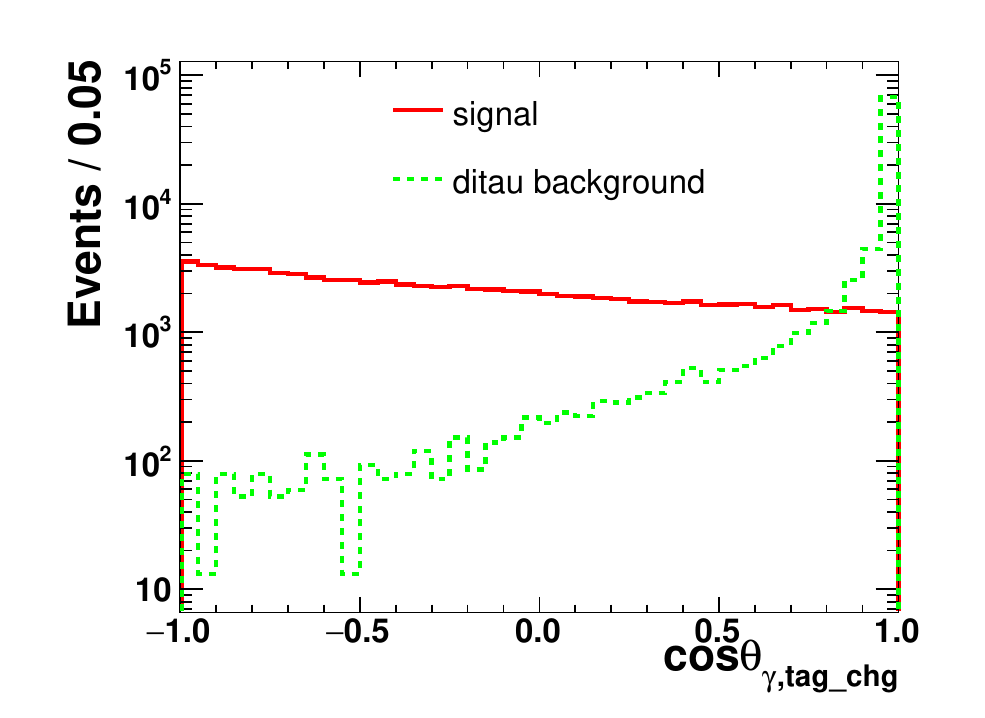}\caption{}\label{subfig:etag_ditau_01_cos_theta_sig_gamma_vs_tag_charged_particle}\end{subfigure}
  \hfill
  \begin{subfigure}{0.49\linewidth}\centering\includegraphics[width = \linewidth]{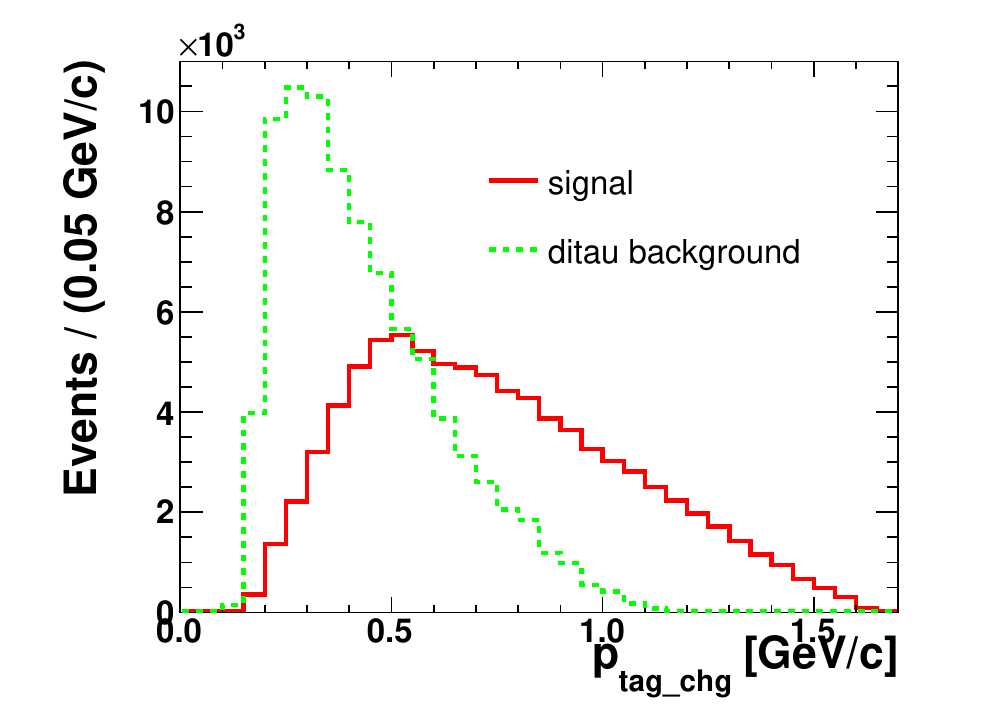}\caption{}\label{subfig:etag_ditau_02_p1_tag_charged_particle}\end{subfigure}
  \begin{subfigure}{0.49\linewidth}\centering\includegraphics[width = \linewidth]{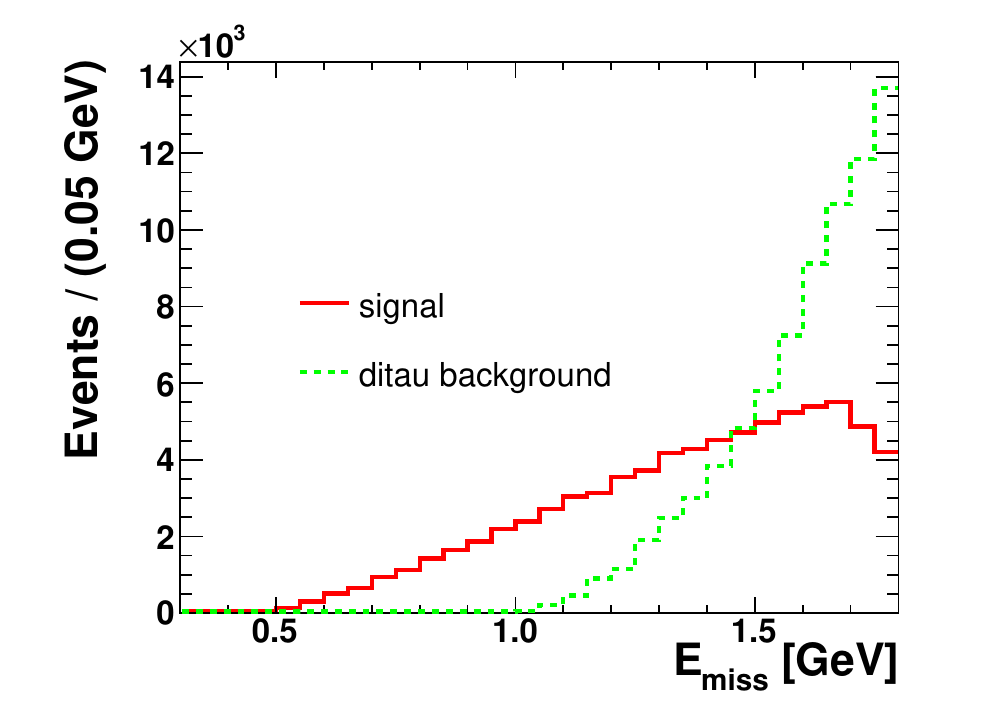}\caption{}\label{subfig:etag_ditau_03_E_miss}\end{subfigure}
  \caption{The comparison of signal (red solid histogram) and background (green dashed histogram) samples in $e^+ \nu_e \bar{\nu}_\tau$ tag mode. \subref{subfig:etag_ditau_01_cos_theta_sig_gamma_vs_tag_charged_particle} the cosine of the angle between signal photon and tag charged track $\cos\theta_{\text{sig\_}\gamma,\text{tag\_charged}}$, \subref{subfig:etag_ditau_02_p1_tag_charged_particle} the momentum of tag charged track $p_{\text{tag\_charged}}$, and \subref{subfig:etag_ditau_03_E_miss} the missing energy $E_{\text{miss}}$ in ditau background.}
  \label{fig:compare_signal_bkg_after_initial_cuts_etag}
\end{figure}

\begin{figure}[htbp]
  \begin{subfigure}{0.49\linewidth}\centering\includegraphics[width = \linewidth]{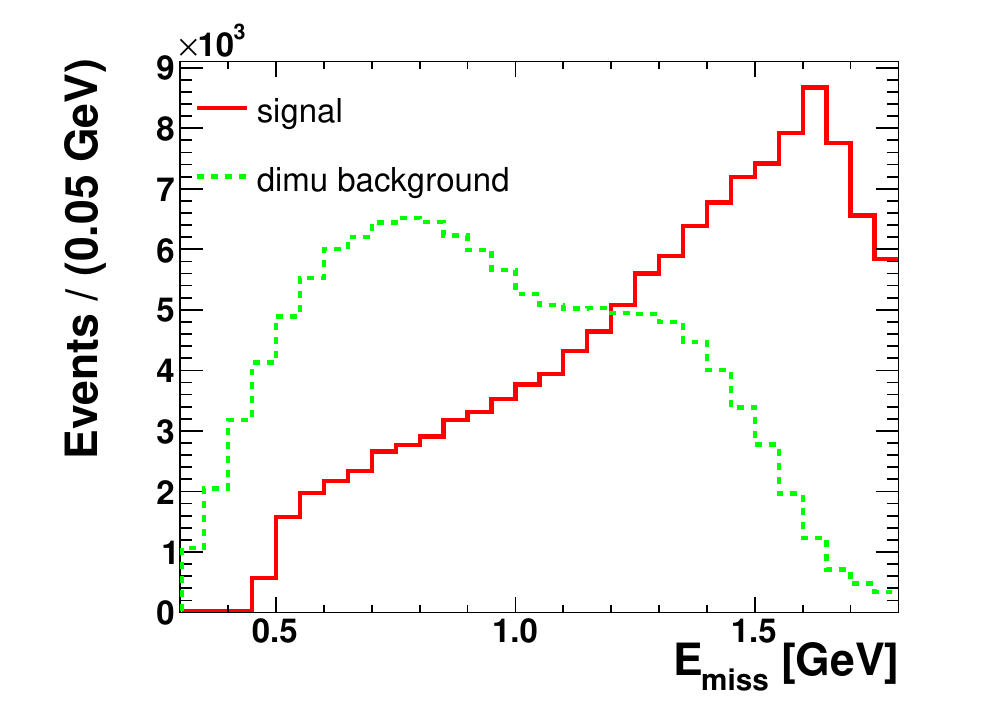}\caption{}\label{subfig:pitag_dimu_01_E_miss}\end{subfigure}
  \hfill
  \begin{subfigure}{0.49\linewidth}\centering\includegraphics[width = \linewidth]{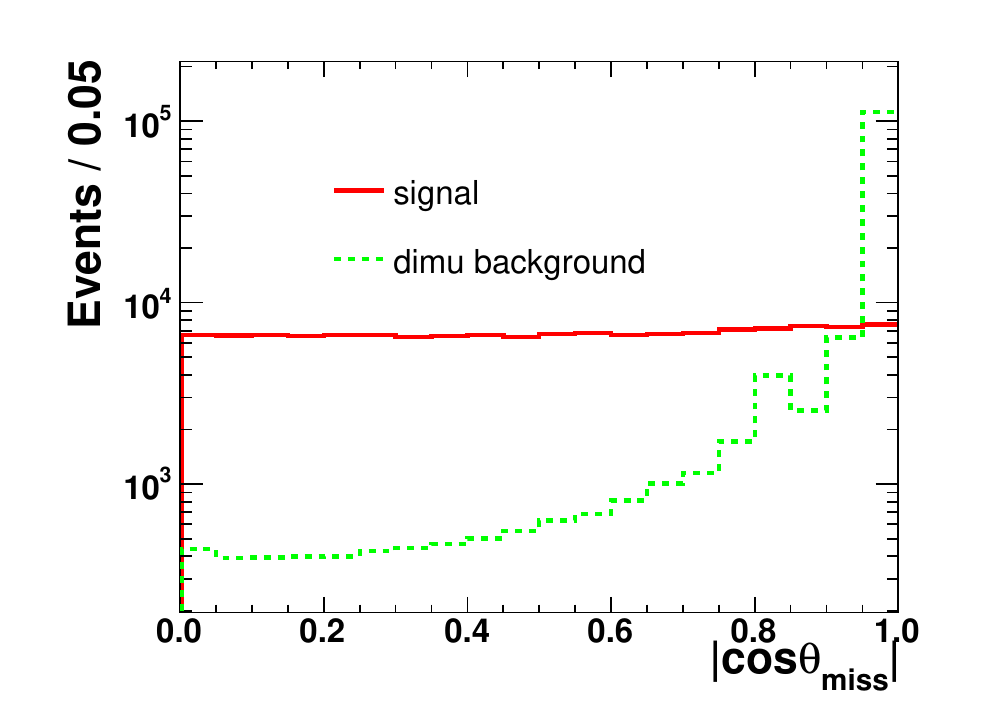}\caption{}\label{subfig:pitag_dimu_02_cos_theta_miss}\end{subfigure}
  \begin{subfigure}{0.49\linewidth}\centering\includegraphics[width = \linewidth]{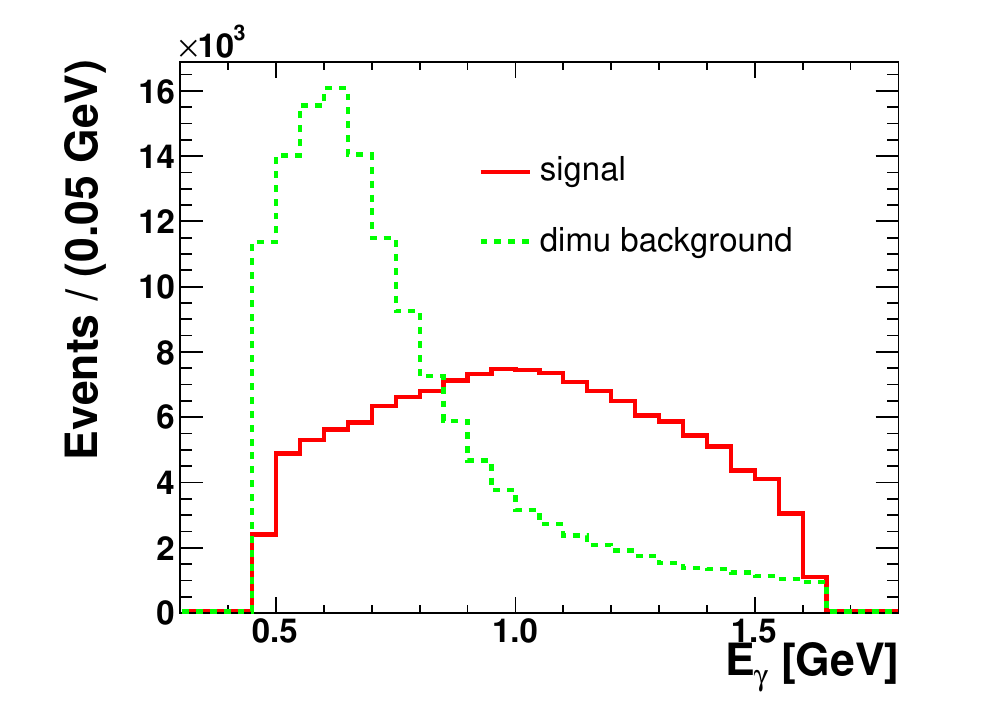}\caption{}\label{subfig:pitag_dimu_04_E_sig_gamma}\end{subfigure}
  \hfill
  \begin{subfigure}{0.49\linewidth}\centering\includegraphics[width = \linewidth]{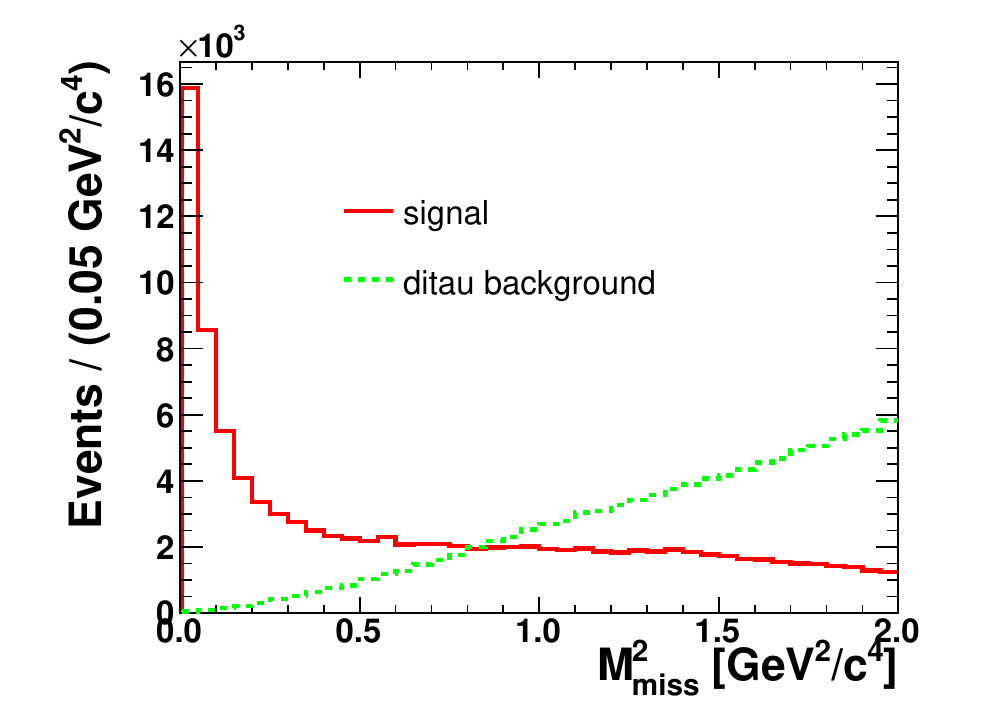}\caption{}\label{subfig:pitag_ditau_M_sq_miss}\end{subfigure}
  \caption{The comparison of signal (red solid histogram) and background (green dashed histogram) samples in $\pi^+ \bar{\nu}_\tau$ tag mode. \subref{subfig:pitag_dimu_01_E_miss} the missing energy $E_{\text{miss}}$, \subref{subfig:pitag_dimu_02_cos_theta_miss} the absolute value of cosine of angle of missing momentum with respect to beam direction $\abs{\cos\theta_{\text{miss}}}$, and \subref{subfig:pitag_dimu_04_E_sig_gamma} the energy of signal photon $E_{\text{sig\_}\gamma}$ in dimu background, and \subref{subfig:pitag_ditau_M_sq_miss} the missing mass squared $M^{2}_{\text{miss}}$ in ditau background.}
  \label{fig:compare_signal_bkg_after_initial_cuts_pitag}
\end{figure}

\begin{figure}[htbp]
  \begin{subfigure}{0.49\linewidth}\centering\includegraphics[width = \linewidth]{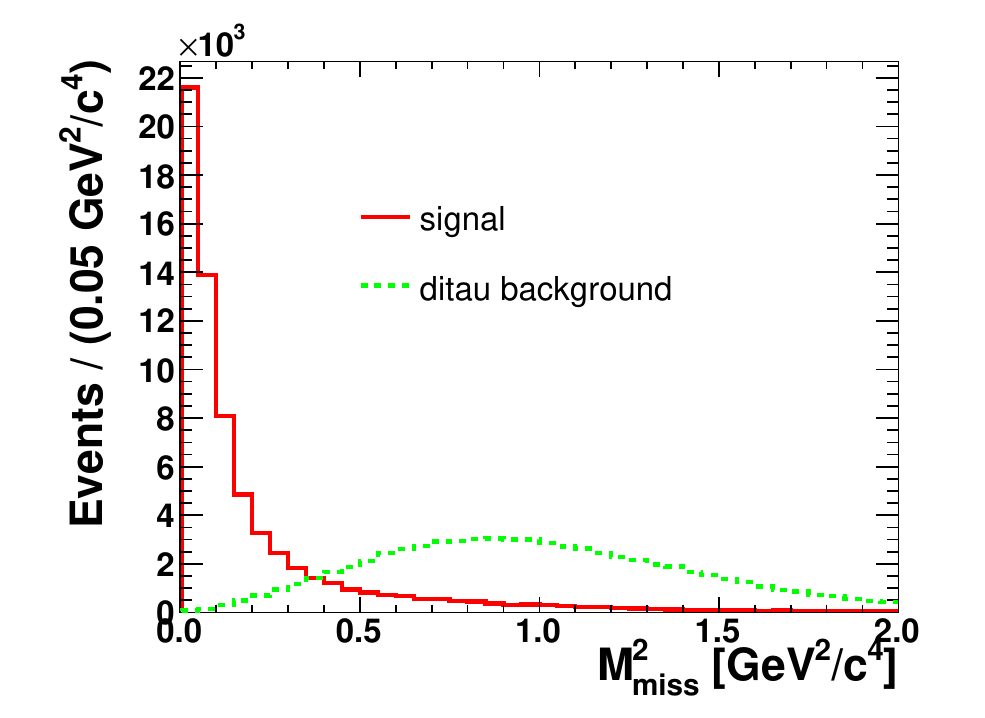}\caption{}\label{subfig:pipi0tag_ditau_01_M_sq_miss}\end{subfigure}
  \hfill
  \begin{subfigure}{0.49\linewidth}\centering\includegraphics[width = \linewidth]{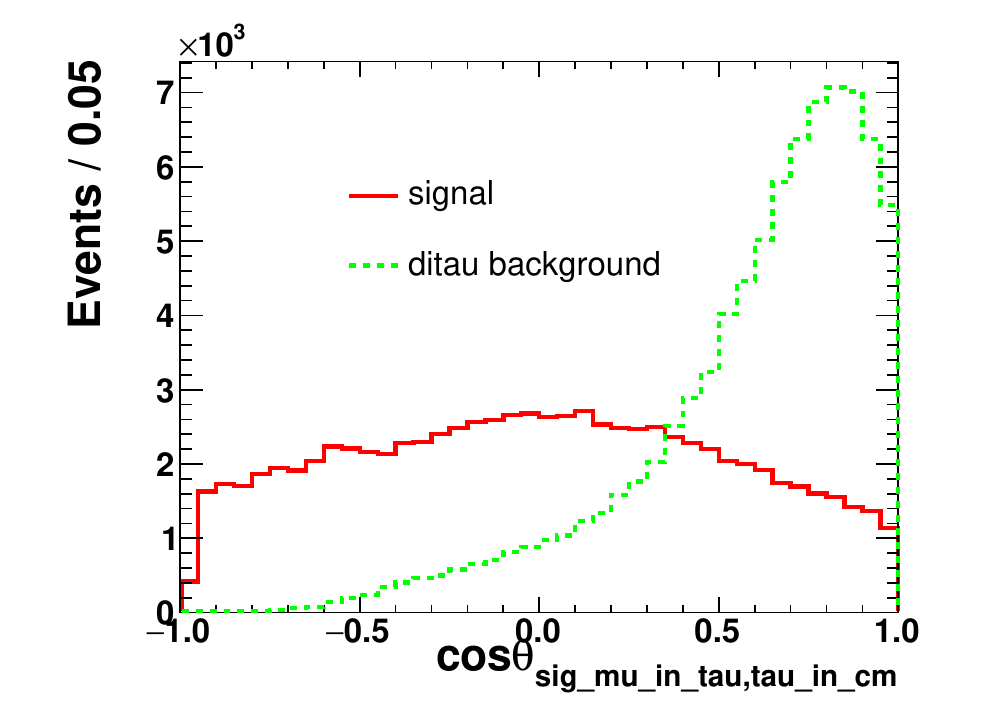}\caption{}\label{subfig:pipi0tag_ditau_02_cos_theta_sig_mu_in_tau_vs_tau_in_cm}\end{subfigure}
  \begin{subfigure}{0.49\linewidth}\centering\includegraphics[width = \linewidth]{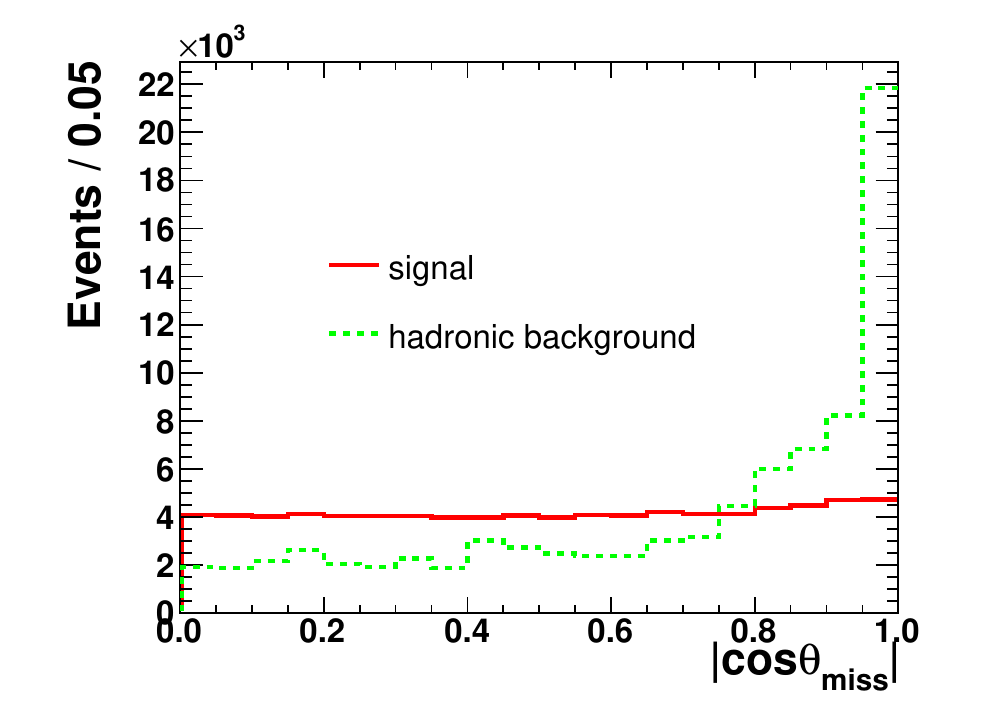}\caption{}\label{subfig:pipi0tag_hadron_cos_theta_miss}\end{subfigure}
  \caption{The comparison of signal (red solid histogram) and background (green dashed histogram) samples in $\pi^+ \pi^0 \bar{\nu}_\tau$ tag mode. \subref{subfig:pipi0tag_ditau_01_M_sq_miss} the missing mass squared $M^{2}_{\text{miss}}$ and \subref{subfig:pipi0tag_ditau_02_cos_theta_sig_mu_in_tau_vs_tau_in_cm} the cosine of the helicity angle of signal muon $\cos\theta_{\text{sig\_mu\_in\_tau},\text{tau\_in\_cm}}$ in ditau background, and \subref{subfig:pipi0tag_hadron_cos_theta_miss} direction of missing momentum $\abs{\cos\theta_{\text{miss}}}$ in hadronic background.}
  \label{fig:compare_signal_bkg_after_initial_cuts_pipi0tag}
\end{figure}

Further event selection criteria are determined based on the background characteristics. For the ditau background in the $e^+ \nu_e \bar{\nu}_\tau$ tag mode, the signal photon is from radiative leptonic decay of tag tau, thus collinear with the tag charged track, as shown in Fig.~\ref{subfig:etag_ditau_01_cos_theta_sig_gamma_vs_tag_charged_particle}. Moreover, the momentum of tag charged track is lowered due to the existence of radiative photon (Fig.~\ref{subfig:etag_ditau_02_p1_tag_charged_particle}). Since neutrinos in the final states are not detected, there will be missing four-momentum defined as the total initial four-momentum subtracted by the four-momenta of all detected final state particles. There are more neutrinos in background than in signal, so the missing energy $E_{\text{miss}}$ in background is higher (Fig.~\ref{subfig:etag_ditau_03_E_miss}), where $E_{\text{miss}}$ is defined as energy component of missing four-momentum. For the radiative dimu background in $\pi^+ \bar{\nu}_\tau$ tag mode, $E_{\text{miss}}$ is lower since there are no neutrinos (Fig.~\ref{subfig:pitag_dimu_01_E_miss}), and the direction of missing momentum accumulates along the beam direction since missing momentum is mainly due to radiative photons which are collinear with beam and can escape in the beam direction which is beyond detector coverage (Fig.~\ref{subfig:pitag_dimu_02_cos_theta_miss}). Furthermore, the energy of the signal photon is lower since it is from radiation (Fig.~\ref{subfig:pitag_dimu_04_E_sig_gamma}). For ditau background in $\pi^+ \bar{\nu}_\tau$ tag mode, in contrast to the signal which has only one neutrino, there are more neutrinos in background, so the missing mass squared $M^{2}_{\text{miss}}$ is around zero for signal while not for background (Fig.~\ref{subfig:pitag_ditau_M_sq_miss}). The $M^{2}_{\text{miss}}$ is defined as the square of invariant mass of the missing four-momentum. For ditau background in $\pi^+ \pi^0 \bar{\nu}_\tau$ tag mode, the $M^{2}_{\text{miss}}$ distribution has similar characteristic with $\pi^+ \bar{\nu}_\tau$ tag mode (Fig.~\ref{subfig:pipi0tag_ditau_01_M_sq_miss}). The distribution of helicity angle of signal muon is also different in signal and background due to different decay dynamics of signal tau (Fig.~\ref{subfig:pipi0tag_ditau_02_cos_theta_sig_mu_in_tau_vs_tau_in_cm}), where the helicity angle is defined as the angle between the direction of signal muon in the signal tau rest frame and the direction of signal tau in center-of-mass frame. For the hadronic background in $\pi^+ \pi^0 \bar{\nu}_\tau$ tag mode, which is mainly $\pi^+\pi^- + (\geq 1)\pi^0$, the missing momentum is due to photons escaping along the beam direction (Fig.~\ref{subfig:pipi0tag_hadron_cos_theta_miss}).

To determine the concrete selection criteria, Punzi significance $\varepsilon/(1.5 + \sqrt{N_{\text{bkg}}})$~\cite{Punzi} is used as the figure of merit, where $\varepsilon$ is signal efficiency and $N_{\text{bkg}}$ is the number of background events. A multidimensional optimization is performed for all the criteria simultaneously. The optimization result depends on the detector performance design, and the result with the best performance is shown here. Table~\ref{tab:result_further_selection} summarizes the further selection criteria and background levels and signal efficiencies before and after further selection. The final background level is suppressed to be only a few with signal efficiency of several percents.

\begin{table}[htbp]
  \centering
  \caption{The result of further event selection. The first column is the tag modes, the second column is the selection criteria, the third and fourth columns are the number of background $N_{\text{bkg}}$ and signal efficiency $\epsilon$ before and after further selection.}
  \label{tab:result_further_selection}
  \resizebox{\linewidth}{!}{
    \begin{tabular}{c|c|c|c}
      \hline
      \hline
      tag mode                                     & selection criteria                                               & $N_{\text{bkg}}$ ($\epsilon$) before             & $N_{\text{bkg}}$ ($\epsilon$) after           \\
      \hline
      \multirow{3}*{$e^+ \nu_e \bar{\nu}_\tau$}    & $\cos\theta_{\text{sig\_}\gamma,\text{tag\_charged}} < -0.2$     & \multirow{3}*{\num{1.5e2} (\SI{2.6}{\percent})}  & \multirow{3}*{\num{0} (\SI{1.1}{\percent})}   \\
                                                   & $p_{\text{tag\_charged}} > \SI{0.5}{GeV/\clight}$                &                                                  &                                               \\
                                                   & $E_{\text{miss}} < \SI{1.7}{GeV}$                                &                                                  &                                               \\
      \hline
      \multirow{4}*{$\pi^+ \bar{\nu}_\tau$}        & $E_{\text{miss}} > \SI{0.7}{GeV}$                                & \multirow{4}*{\num{1.4e4} (\SI{4.0}{\percent})}  & \multirow{4}*{\num{0.3} (\SI{0.5}{\percent})} \\
                                                   & $\abs{\cos\theta_{\text{miss}}} < 0.6$                           &                                                  &                                               \\
                                                   & $E_{\text{sig\_}\gamma} > \SI{0.8}{GeV}$                         &                                                  &                                               \\
                                                   & $M^{2}_{\text{miss}} < \SI{0.050}{GeV^{2}/\clight^{4}}$          &                                                  &                                               \\
      \hline
      \multirow{3}*{$\pi^+ \pi^0 \bar{\nu}_\tau$}  & $M^{2}_{\text{miss}} < \SI{0.075}{GeV^{2}/\clight^{4}}$          & \multirow{3}*{\num{1.2e3} (\SI{2.6}{\percent})}  & \multirow{3}*{\num{1.2} (\SI{1.5}{\percent})} \\
                                                   & $\cos\theta_{\text{sig\_mu\_in\_tau},\text{tau\_in\_cm}} < 0.8$  &                                                  &                                               \\
                                                   & $\abs{\cos\theta_{\text{miss}}} < 0.9$                           &                                                  &                                               \\
      \hline
      total                                        &                                                                  & \num{1.6e4} (\SI{9.2}{\percent})                 & \num{1.5} (\SI{3.1}{\percent})                \\
      \hline
      \hline
    \end{tabular}
  }
\end{table}

\begin{figure}[htbp]
  \centering
  \includegraphics[width = \linewidth]{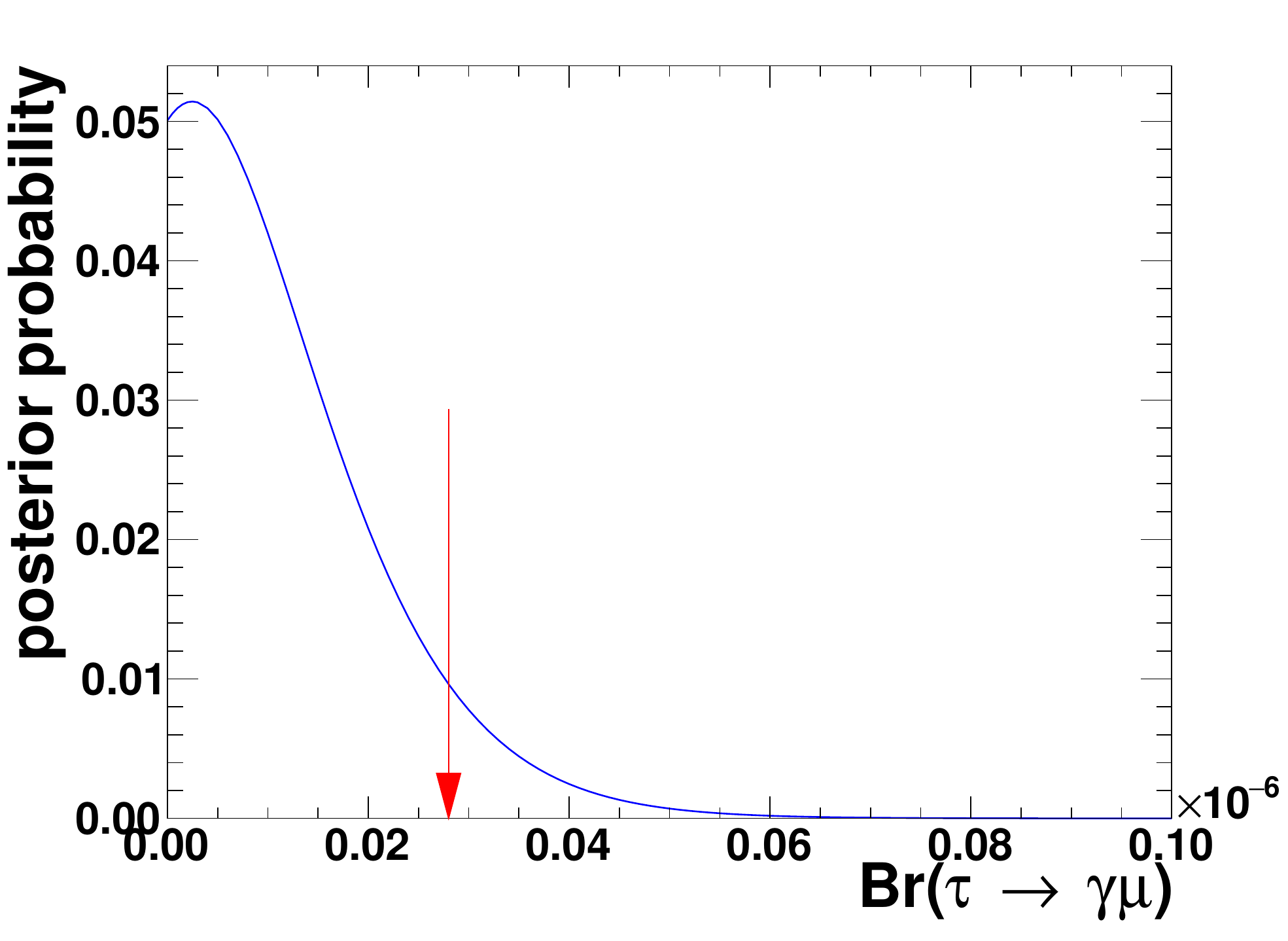}
  \caption{Determination of UL of BF. The line shows the posterior probability distribution and the arrow marks the UL at \SI{90}{\percent} C.L.}
  \label{fig:UL}
\end{figure}

A Bayesian-based maximum likelihood estimator, extended from the profile likelihood approach~\cite{UL_method}, is used to determine the UL on BF of $\tau \to \gamma\mu$ with statistical fluctuations taking into account. The likelihood is constructed as
\begin{equation}
  \label{eq:likelihood}
  \begin{split}
    \mathcal{L} &= \text{Poisson}(N_{\text{obs}}, 2N_{\tau^+\tau^-} \times \mathcal{B} \times \epsilon + \sum_{i}N_{\text{bkg},i}) \\
    &\times \prod_{i}\text{Poisson}(N_{\text{bkg},i}^{\text{obs}}, N_{\text{bkg},i}/f_i),
  \end{split}
\end{equation}
where $N_{\text{obs}}$ is the observed number of events, $N_{\tau^+\tau^-}$ is the number of tau pairs, $\mathcal{B}$ is BF of $\tau \to \gamma\mu$ and is the parameter of interest, $\epsilon$ is signal efficiency, $N_{\text{bkg},i}$ with $i$ runs over all the background samples are the \emph{true} values of background levels which are nuisance parameters, $N_{\text{bkg},i}^{\text{obs}}$ and $f_i$ are the observed number of events and scale factors for each background samples. The likelihood is then taken as probability distribution of parameters, and the posterior distribution of BF is obtained by integrating over nuisance parameters. Finally, the UL on BF is determined by integrating the posterior distribution, as shown in Fig.~\ref{fig:UL}. The sensitivity of BF is defined as the UL that can be achieved assuming there are no signals. Pseudo-experiments are generated assuming background-only, and the median, the $\pm1\sigma$ and $\pm2\sigma$ quantiles on the distribution of UL are taken as the expectation and confidence interval of the sensitivity~\cite{expected_limit}. With a total luminosity of \SI{1}{ab^{-1}}, the sensitivity is estimated to be at the level of \num{e-8}.

A full systematic uncertainty evaluation which requires both experimental data and full MC simulation is not possible at this stage, so it will be qualitatively discussed. Referring to Eq.~\ref{eq:likelihood}, the possible sources of systematic uncertainties include the number of tau pairs, the event selection efficiency, and the estimation of background. The uncertainty of number of tau pairs comes from the determination of luminosity and cross-section of $e^+e^- \to \tau^+\tau^-$. For efficiency, the statistical uncertainty can be negligible with large MC samples; the uncertainty of tracking and PID of tracks and reconstruction of neutral pions, which is evaluated with difference between data and MC, can be studied with pure and high statistical control samples; uncertainties related with other selection criteria can be evaluated by control samples or varying the criteria and performing the Barlow test~\cite{Barlow_test}; for the modeling of $\tau \to \gamma\mu$ decay, although the LFV interaction structure in unknown, the uncertainty can be estimated by assuming extreme cases such as pure $V-A$ and $V+A$ forms. The uncertainty of background estimation can be evaluated using control samples of data, such as from sidebands, and verifying that the estimated background is consistent with that in real data. The total systematic uncertainties at STCF are expected to be at the level of several percent or less, which only have a minor impact on the sensitivity.

\section{Optimization of detector performance}
\label{sec:optimization}
The performance of STCF detector is tunable in the fast simulation, and the sensitivity of $\tau \to \gamma\mu$ under different performances are studied to guide the design of the detector.

The detector performance properties that are crucial to this analysis are determined based on the main background channels where photons and muons are misidentified as signals. One of the main origins of the signal muon is misidentification of pion, while the signal photon is misidentified from a photon with other origins. So, both pion/muon separation capability and photon detection resolution are relevant to this analysis. Better pion/muon separation capability can efficiently suppress background caused by pion/muon misidentification, and better photon detection resolution will improve the resolution of signal region thus exclude more background. With fast simulation, three kinds of detector responses can be studied: pion/muon separation, photon energy resolution and photon position resolution. Considering the feasibility of particular detector designs, a set of different values is assumed for each of the performance properties, from conservative to aggressive. The best detector performance is taken as benchmark, and the dependence of the sensitivity upon each performance property is checked by fixing the other properties and only varying the one under study. The benchmark result is sensitivity of \num{2.8e-8} with \SI{1}{ab^{-1}} luminosity under the detector performance of \SI{1}{\percent} in pion/muon misidentification rate and \SI{3}{mm} and \SI{2}{\percent} in photon position and energy resolutions. For each performance setting, the analysis is performed individually to reach the best sensitivity.

\begin{figure}[htbp]
  \centering
  \includegraphics[width = \linewidth]{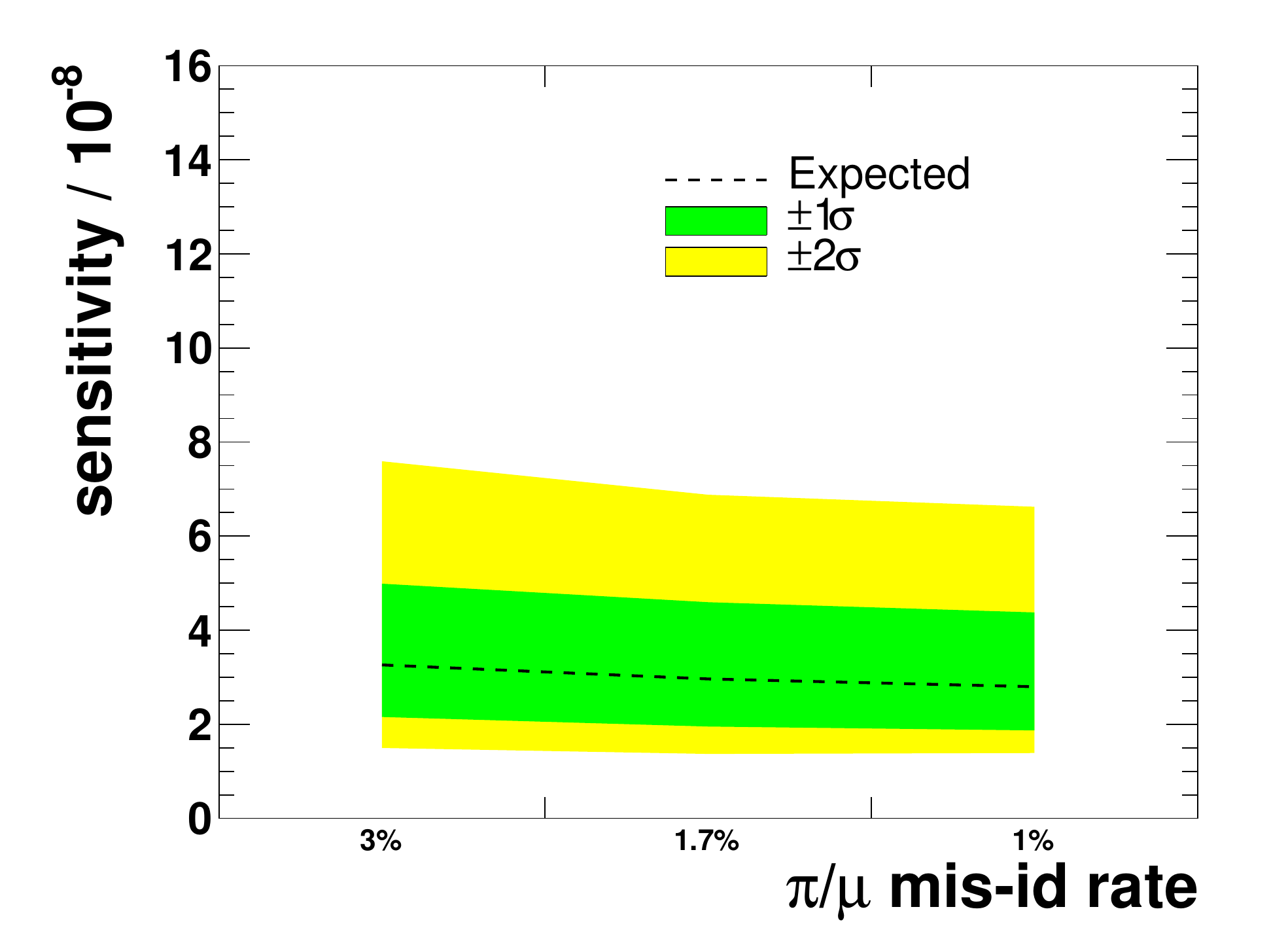}
  \caption{Result for optimization of pion/muon separation capability. The dashed line is the expected sensitivity, the green and yellow bands are the $1\sigma$ and $2\sigma$ confidence intervals.}
  \label{fig:detector_optimization_result_pi_mu_separation}
\end{figure}

\paragraph{pion/muon separation}
On the one hand, better pion/muon separation can suppress the background caused by pion misidentified as signal muon. On the other hand, better separation means tighter muon selection, which will cause lower signal efficiency. Three levels of pion/muon separation capability is assumed with overall pion to muon misidentification rates of \SI{3}{\percent}, \SI{1.7}{\percent} and \SI{1}{\percent}, which corresponding to the muon identification efficiencies of \SI{97}{\percent}, \SI{92}{\percent} and \SI{85}{\percent} at a momentum of \SI{1}{GeV/\clight}, respectively. Table~\ref{tab:optimization_pi_mu_misid} summarizes the efficiency of muon identification in signal sample and the sensitivity of $\tau \to \gamma\mu$ with respect to pion/muon misidentification rate. The sensitivity together with its confidence intervals are also shown in Fig.~\ref{fig:detector_optimization_result_pi_mu_separation}. The result shows that the sensitivity improves with better pion/muon separation, but the improvement is small. This indicates that a trade-off between the muon efficiency and the pion/muon separation capability should be carefully considered.

\begin{table}[htbp]
  \centering
  \caption{Result of optimization for pion/muon separation. The first column shows the levels of pion/muon separation capability of the detector. The second and third columns show the muon identification efficiency and sensitivity on BF under different detector performance, separately.}
  \label{tab:optimization_pi_mu_misid}
  \begin{tabular}{cccc}
    \hline
    \hline
    pion/muon mis-id rate  & muon PID eff.        & sensitivity/\num{e-8} \\
    \hline
    \SI{3}{\percent}       & \SI{80.6}{\percent}  & \num{3.3}             \\
    \SI{1.7}{\percent}     & \SI{65.4}{\percent}  & \num{3.0}             \\
    \SI{1}{\percent}       & \SI{50.3}{\percent}  & \num{2.8}             \\
    \hline
    \hline
  \end{tabular}
\end{table}

\begin{figure}[htbp]
  \centering
  \includegraphics[width = 0.49\linewidth]{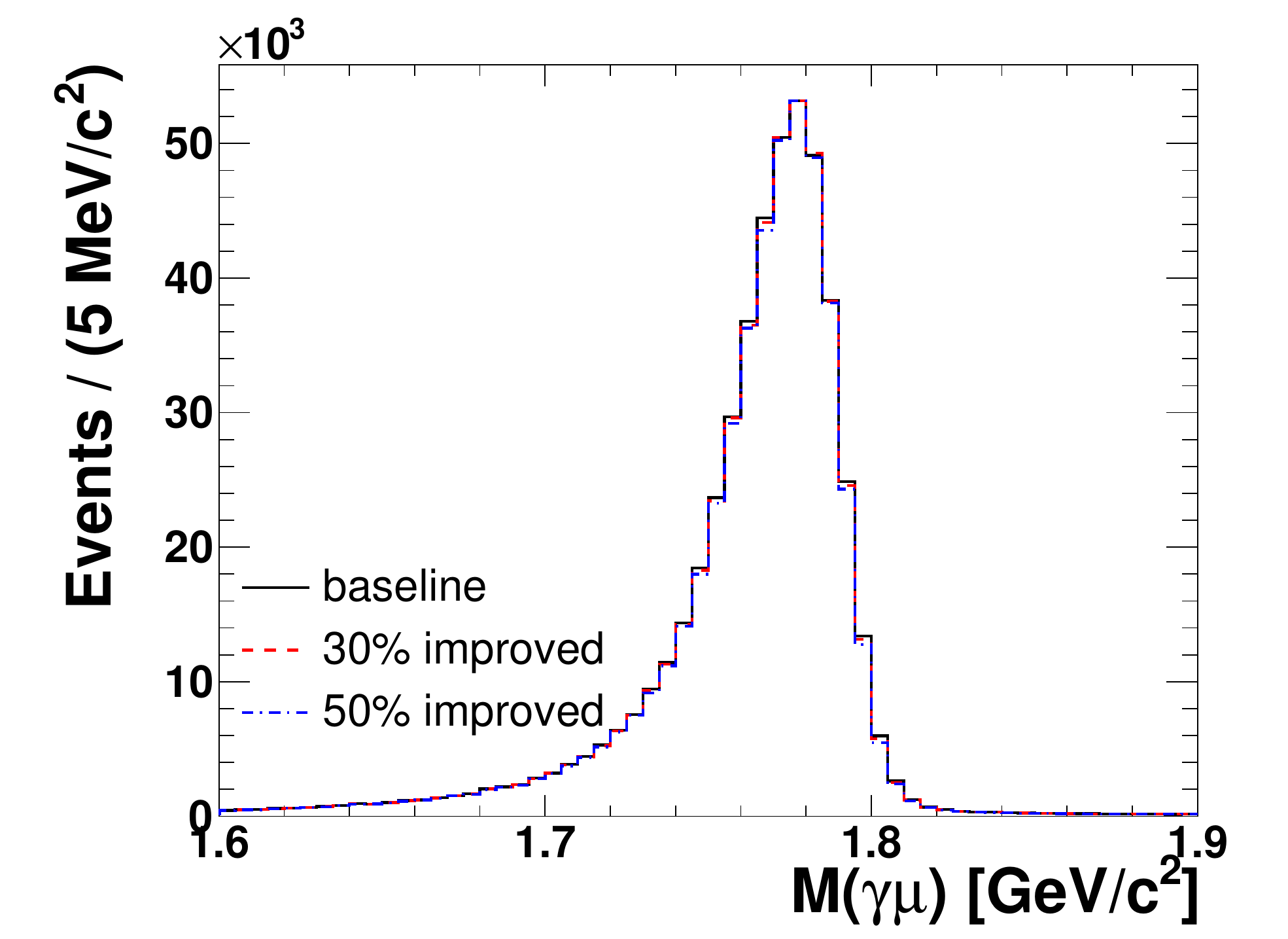}
  \includegraphics[width = 0.49\linewidth]{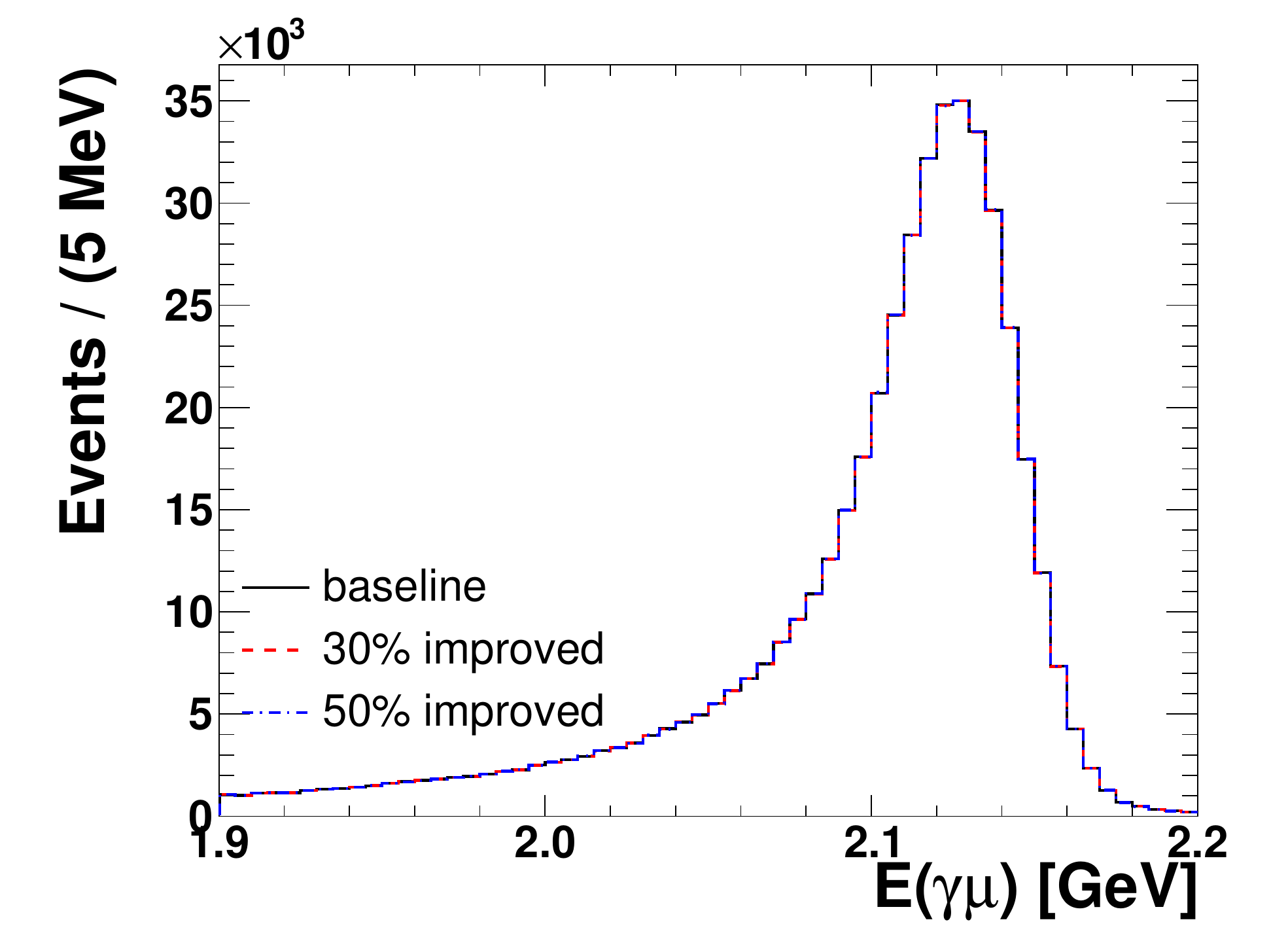}
  \caption{Signal resolution under different photon position resolution performance. The left and right figure shows the invariant mass and total energy of signal photon and muon, separately. The black solid line, the red dashed line and the blue dash-dotted line shows the result under baseline, \SI{30}{\percent} and \SI{50}{\percent} improved photon position resolution, separately.}
  \label{fig:signal_resolution_vs_photon_position_resolution}
\end{figure}

\begin{figure}[htbp]
  \centering
  \includegraphics[width = \linewidth]{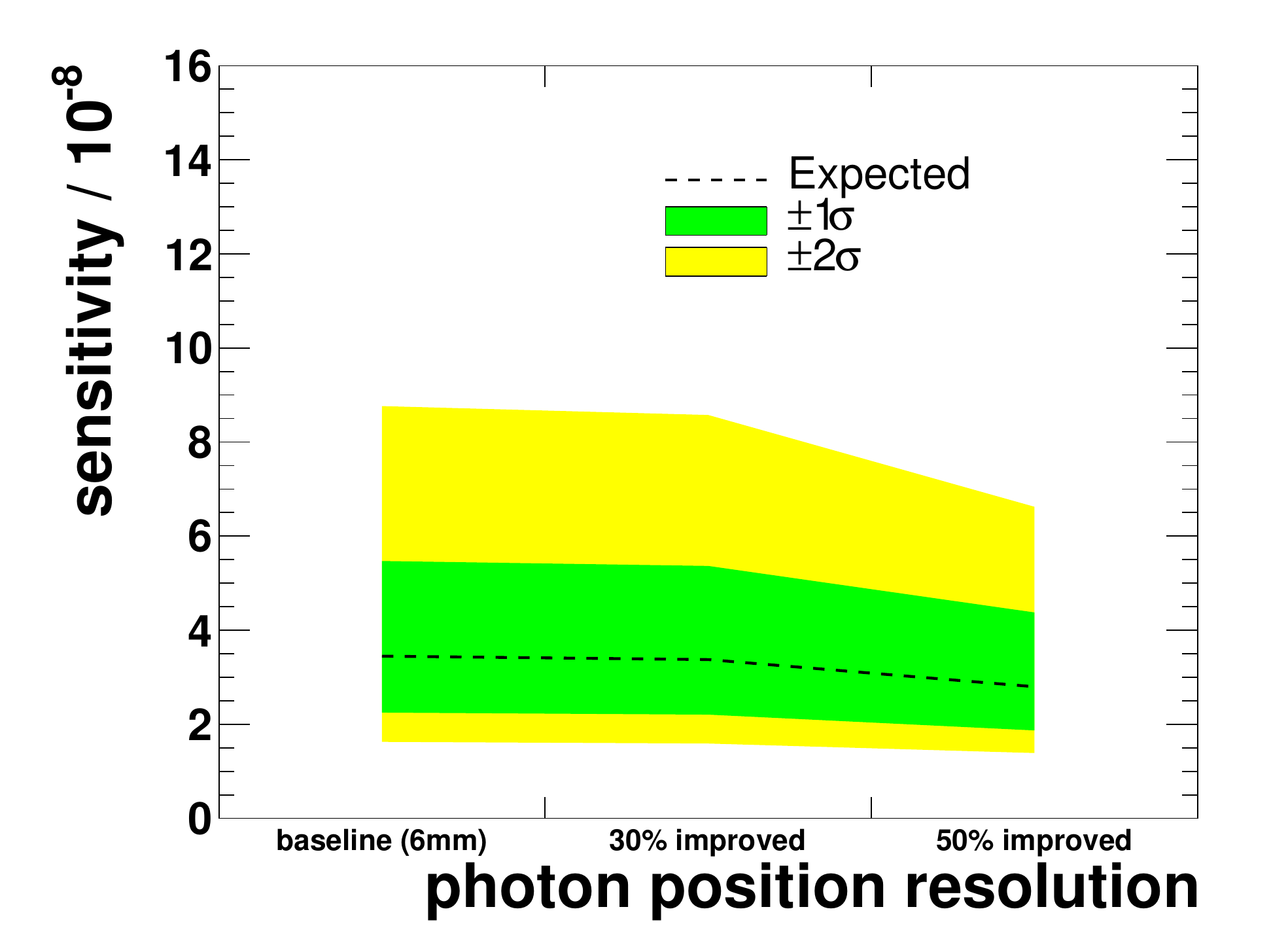}
  \caption{Result for optimization of photon position resolution.}
  \label{fig:detector_optimization_result_photon_position_resolution}
\end{figure}

\paragraph{position resolution for photon}
Better photon position resolution will result in better signal region resolution, thus improve efficiency and suppress background. Besides, better photon resolution will result in higher tag accuracy and better separation of noise from real photons. The baseline for photon position resolution is \SI{6}{mm} and an improvement of \SI{30}{\percent} and \SI{50}{\percent} is assumed for optimization. The signal resolution under different photon position resolutions is shown in Fig.~\ref{fig:signal_resolution_vs_photon_position_resolution}, and the sensitivity is summarized in Table~\ref{tab:optimization_photon_position_resolution} and Fig.~\ref{fig:detector_optimization_result_photon_position_resolution}. It is shown that better photon position resolution will result in better sensitivity, but the influence is rather small, which is because the baseline resolution is already quite good.

\begin{table}[htbp]
  \centering
  \caption{Result of optimization for photon position resolution. The first column shows the levels of photon position resolution performance of the detector. The second and third columns show the signal efficiency and sensitivity on BF under different detector performance.}
  \label{tab:optimization_photon_position_resolution}
  \begin{tabular}{cccc}
    \hline
    \hline
    resolution                  & $\epsilon$          & sensitivity/\num{e-8} \\
    \hline
    baseline                    & \SI{3.0}{\percent}  & 3.4                   \\
    \SI{30}{\percent} improved  & \SI{3.1}{\percent}  & 3.4                   \\
    \SI{50}{\percent} improved  & \SI{3.1}{\percent}  & 2.8                   \\
    \hline
    \hline
  \end{tabular}
\end{table}

\begin{figure}[htbp]
  \centering
  \includegraphics[width = 0.49\linewidth]{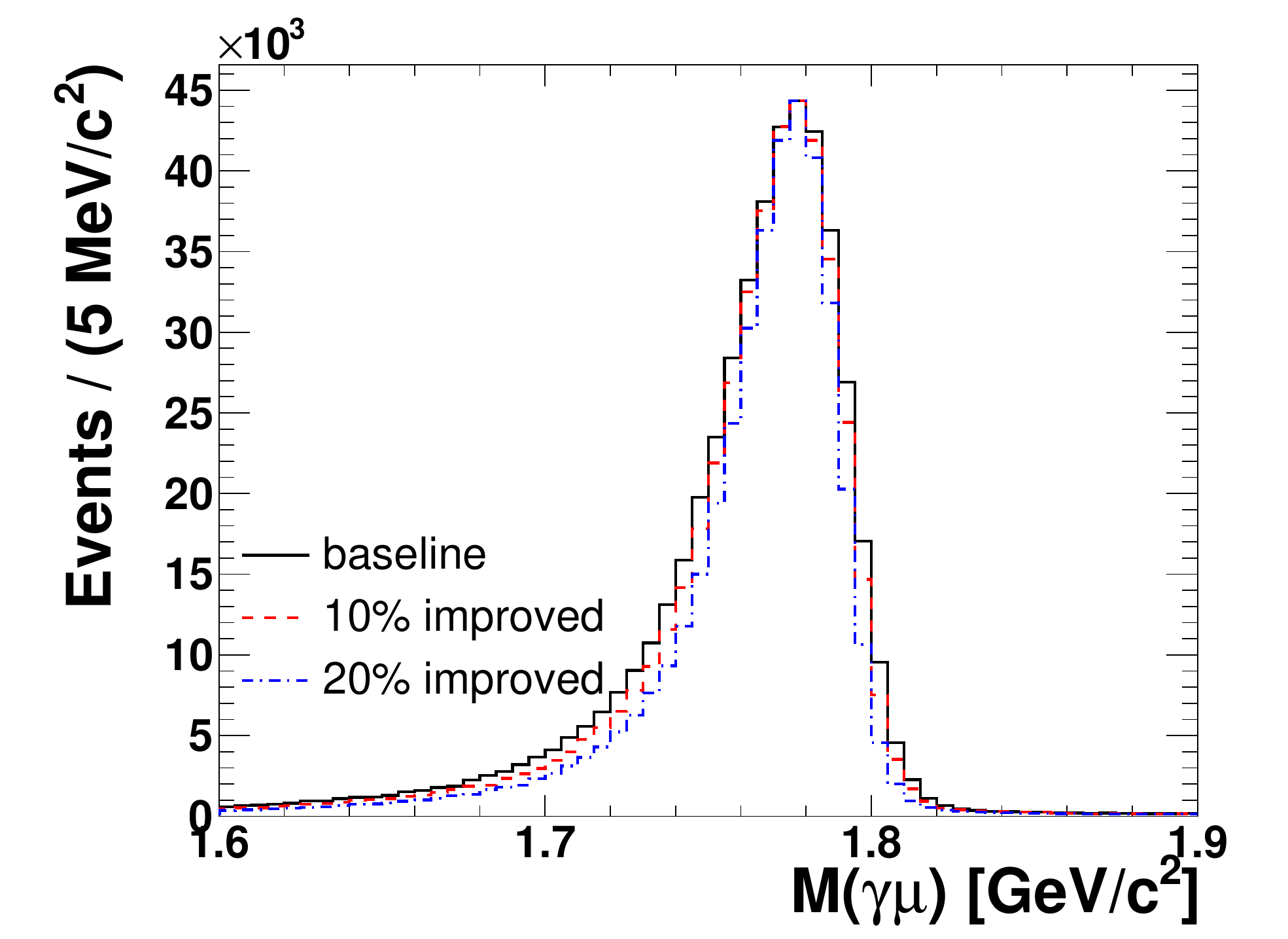}
  \includegraphics[width = 0.49\linewidth]{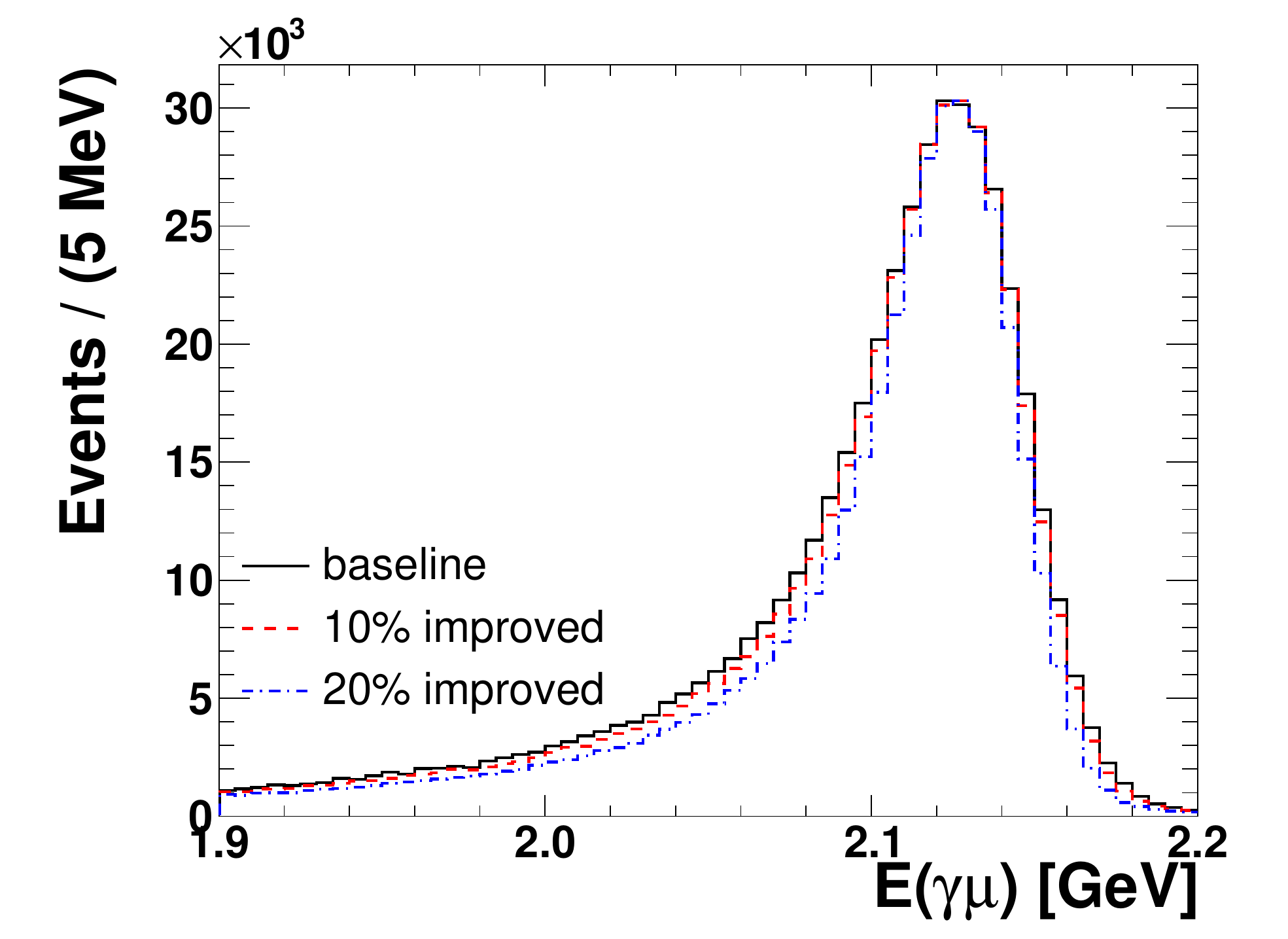}
  \caption{Signal resolution under different photon energy resolution performance. The left and right figure shows the invariant mass and total energy of signal photon and muon, separately. The black solid line, the red dashed line and the blue dash-dotted line shows the result under baseline, \SI{10}{\percent} and \SI{20}{\percent} improved photon energy resolution, separately.}
  \label{fig:signal_resolution_vs_photon_energy_resolution}
\end{figure}

\begin{figure}[htbp]
  \centering
  \includegraphics[width = \linewidth]{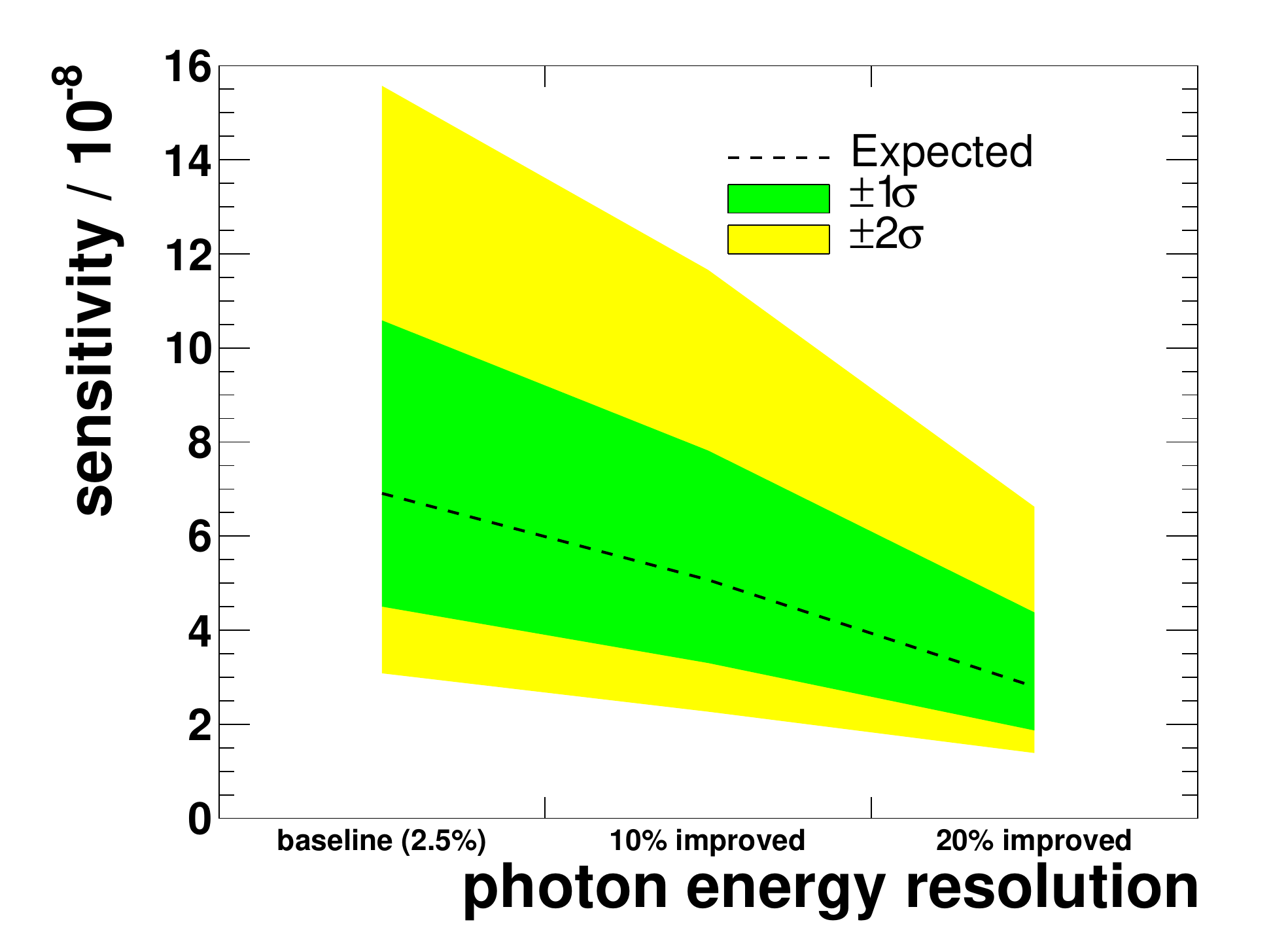}
  \caption{Result for optimization of photon energy resolution.}
  \label{fig:detector_optimization_result_photon_energy_resolution}
\end{figure}

\paragraph{energy resolution for photon}
Similar to photon position resolution, better photon energy resolution will also result in better sensitivity. The baseline for photon energy resolution is \SI{2.5}{\percent} at \SI{1}{GeV} and an improvement of \SI{10}{\percent} and \SI{20}{\percent} is assumed for optimization. The signal resolution under different photon energy resolutions is shown in Fig.~\ref{fig:signal_resolution_vs_photon_energy_resolution}, and the sensitivity is summarized in Table~\ref{tab:optimization_photon_energy_resolution} and Fig.~\ref{fig:detector_optimization_result_photon_energy_resolution}. It is shown that better photon energy resolution will result in better sensitivity.

\begin{table}[htbp]
  \centering
  \caption{Result of optimization for photon energy resolution. The first column shows the levels of photon energy resolution performance of the detector. The second and third columns show the signal efficiency and sensitivity on BF under different detector performance.}
  \label{tab:optimization_photon_energy_resolution}
  \begin{tabular}{cccc}
    \hline
    \hline
    resolution                  & $\epsilon$          & sensitivity/\num{e-8} \\
    \hline
    baseline                    & \SI{2.9}{\percent}  & 6.9                   \\
    \SI{10}{\percent} improved  & \SI{3.0}{\percent}  & 5.1                   \\
    \SI{20}{\percent} improved  & \SI{3.1}{\percent}  & 2.8                   \\
    \hline
    \hline
  \end{tabular}
\end{table}

\section{Summary and discussion}
The sensitivity on the cLFV process $\tau \to \gamma \mu$ at the Super $\tau$-Charm Facility is studied based on \SI{1}{ab^{-1}} MC samples, which corresponds to the one-year integrated luminosity of STCF. The sensitivity is expected to be at the level of \num{e-8}. The optimization of detector performance is also studied in order to get the best sensitivity, and the result shows that the improvement of each of the three performance properties concerned in this analysis, namely pion/muon separation capability, position and energy resolution of photon, can all result in better sensitivity, though the importance of them is different. With ideal detector performance of \SI{1}{\percent} in pion/muon misidentification rate and \SI{3}{mm} and \SI{2}{\percent} in photon position and energy resolutions, the best sensitivity of \num{2.8e-8} at \SI{90}{\percent} confidence level is achieved. Since background-free can not be achieved, the sensitivity is expected to scale with the square root of the luminosity, and could reach \num{8.8e-9} with ten-year of data taking, which is about one order of magnitude improvement upon the current best result.

\begin{acknowledgements}
  We thank the Hefei Comprehensive National Science Center for their strong support. This work is supported by the Joint Large-Scale Scientific Facility Funds of the National Natural Science Foundation of China and Chinese Academy of Sciences (CAS) under Contract No. U1832207, the National Key R\&D Program of China under Contracts No. 2020YFA0406400, and the international partnership program of the CAS Grant No. 211134KYSB20200057.
\end{acknowledgements}

\printbibliography

\end{document}